\documentclass[twocolumn,showpacs,preprintnumbers,amsmath,amssymb]{revtex4}
\usepackage{graphicx}
\usepackage{textcomp}

\begin{document}
\title{Optical analogue of population trapping in the continuum: classical and quantum interference effects}
  \normalsize
\author{Stefano Longhi 
}
\address{Dipartimento di Fisica and Istituto di Fotonica e
Nanotecnologie del CNR, Politecnico di Milano, Piazza L. da Vinci
32, I-20133 Milano, Italy}

%
\bigskip
\begin{abstract}
\noindent
 A quantum theory of light propagation in
two optical channel waveguides tunnelling-coupled to a common
continuum of modes (such as those of a slab waveguide) is
presented, and classical and quantum interference effects are
investigated. For classical light, the photonic system realizes an
optical analogue of coherent population trapping in the continuum
encountered in atomic physics, where destructive interference
between different light leakage channels leads to the appearance
of a trapped state embedded in the continuum. For nonclassical
light, two-photon interference effects are predicted, such as the
tendency of photon pairs to bunch when decaying into the
continuum.
\end{abstract}


\maketitle

\section{Introduction}
It is well established that quantum mechanics of a
non-relativistic particle and paraxial wave optics in dielectric
media shear strong formal similarities (see, for instance,
\cite{Krivoshlykov,Marte97,Dragoman02}). Owing to such
similarities, the temporal evolution of a quantum particle (e.g.
an electron in an atom or in a crystal) can be mimicked by means
of light propagation in suitably-designed photonic structures.
Quantum-optical analogies have seen in recent years a renewed and
increasing interest, both theoretically and experimentally, mainly
motivated by the possibility offered by optics to visualize at a
macroscopic level certain coherent phenomena, originally proposed
for quantum systems, which may be of difficult access or of
controversial interpretation in the quantum  context (see, e.g.,
\cite{Dragoman03,Longhi09} and references therein). In particular,
recent works theoretically proposed and experimentally
demonstrated the optical analogues of some important coherent
phenomena encountered in atomic and molecular physics, such as
coherent population transfer \cite{Kenis01,Longhi07},
electromagnetically-induced transparency \cite{EIT}, and
population trapping in the continuum
\cite{Longhi08PRA,Longhi08JMO,Marinica08}. The latter effect has
been extensively investigated in atomic physics in connection, for
instance, with the problem of autoionization of an atom by
ultraviolet radiation or in laser-induced continuum structures
(see \cite{Coleman82,Knight90} and references therein). In the
simplest case, population trapping in the continuum describes
decay suppression of two discrete states coupled to a common and
unstructured continuum: as a single bound state coupled to the
continuum decays in an irreversible way and population is
transferred into the continuum, under certain conditions coupling
of the continuum with the additional other bound state may
partially or totally suppress the decay of states owing to a
destructive interference effect which was first studied by Fano
for radiative transitions to autoionizing states in atoms
\cite{Fano}. Fano-like resonances in photonic systems have been
considered in several recent works as well
\cite{scattering1,scattering2}, with a main focus on the
scattering (transmission) properties of the structures. These
previous studies in optics have mostly considered propagation of
classical light in coupled guiding structures, disregarding the
quantum nature of light. For classical light, destructive
interference among different tunneling paths is responsible for
the existence of bound states in the continuum, similarly to Fano
interference in atomic physics. However, when few photons or
nonclassical beams are used to excite the photonic structures,
light propagation may show interference effects of quantum nature.
Since long time, coupled waveguides are known to behave similarly
to beam splitters (see, for instance, \cite{Lai91}) and to show
strictly quantum features when nonclassical light propagates
through them, such as two-photon Hong-Ou-Mandel quantum
interference originally demonstrated for beam splitters
\cite{Hong87}. Like beam splitters \cite{Campos89}, coupled
waveguides offer the possibility to transfer nonclassical
properties of light and to generate entangled states
\cite{Lai91,Kim02}. With recent technological advances in the
realization of high-quality low-loss integrated photonic
structures and nonclassical light sources, such possibilities are
nowadays realities. High-fidelity Hong-Ou-Mandel quantum
interference and integrated optical realizations of other key
quantum photonic circuits have been recently demonstrated in
silicon-based waveguide couplers \cite{Politi08}. In another
recent experiment, Bromberg et al. \cite{Bromberg08} showed
nontrivial photon correlations in coupled waveguide arrays and
observed them using classical intensity correlation measurements.
The possibility offered by experimentalists to test quantum
aspects of light in integrated optic networks motivates to extend
previous studies of quantum-optical analogies from the classical
to the quantum level, highlighting interference effects of purely
quantum nature. For instance, a recent theoretical study of
optical Bloch oscillations and Zener tunneling in optical lattices
\cite{LonghiPRL08} showed that propagation of nonclassical light
in the lattice may reveal certain
particle-like aspects of light and quantum interference phenomena.\\
In this work we investigate classical and quantum interference
effects of light in an optical analogue of coherent population
trapping, based on two optical channel waveguides side coupled to
a common slab waveguide recently proposed in
Ref.\cite{Longhi08PRA}. For classical light (coherent states),
such system realizes in optics a simple analogue of destructive
Fano interference with suppression of light leakage from the
channel waveguides (analogous to two discrete states) into the
slab waveguide (the analogue of the continuum). When the
description of light propagation is made at the quantum level,
photon pair excitation reveals strictly quantum features, such as
the tendency of photons to bunch when decaying into the continuum.

\section{The optical analogue of population trapping in the continuum: basic model
and classic wave optics description}
\subsection{The model}
Let us consider propagation of quasi-monochromatic and paraxial
light wave packets at carrier frequency $\omega=2 \pi c_0/
\lambda$ in a weakling guiding dielectric structure, with optical
axis $z$ and refractive index profile $n(x,y)$, composed by two
straight and parallel single-mode and equal channel waveguides
W$_1$ and W$_2$, side-coupled to a slab waveguide S as shown in
Fig.1(a). In the paraxial, weak guidance and quasi-monochromatic
approximations, the electric field can be written as $E(x,y,z,t)
\propto \psi(x,y,z,t) \exp(i \beta z-i\omega
t)+\psi^{\dag}(x,y,z,t) \exp(-i \beta z+i\omega t)$, where the
envelope $\psi$ varies slowly with respect to $z$ and $t$ over one
wavelength $\lambda$ and one optical cycle $2 \pi / \omega$.
Neglecting nonlinearities and group-velocity dispersion, the slow
evolution of the electric field envelope $\psi(x,y,z,t)$ along the
paraxial $z$ direction is governed by the scalar wave equation
\begin{equation}
i \left( \psi_z+ \frac{1}{v_g} \psi_t \right)=- \frac{1}{2 \beta}
(\psi_{xx}+\psi_{yy})+V(x,y) \psi,
\end{equation}
where $\beta=(\omega / c_0) n_s$ is the reference propagation
constant, $n_s$ is the substrate refractive index, $v_g=(d \beta /
d \omega)^{-1}$ is the group velocity of light, and $V(x,y)=\beta
[n_s-n(x,y)]/n_s$. The electric field envelope $\psi$ has been
normalized such that the cycle-averaged total energy of the
electromagnetic field (see, e.g., \cite{Kennedy88})  is given by
\begin{equation}
U \simeq \frac{\hbar \omega } {v_g} \int dx dy dz \psi^\dag \psi =
\hbar \omega \int dx dy dt \psi^\dag \psi.
\end{equation}
At the entrance plane $z=0$, light is typically injected into
either one, or in both, the channel waveguides W$_1$ and W$_2$, in
the form of either monochromatic waves or wave packets with
spatial profiles tailored to match their fundamental modes. Owing
to evanescent coupling with the slab waveguide S, light leakage
into the continuous set of modes of the slab is generally found,
however for certain geometric settings a trapping state may exist,
corresponding to destructive interference between different
tunneling paths into the continuum \cite{Longhi08PRA}. At a
classical level, light propagation in the waveguides is well
captured in the framework of a coupled mode equation approach, as
previously shown in Ref.\cite{Longhi08PRA}. After introduction of
the spectral decomposition $\psi(x,y,z,t)=(2 \pi)^{-1/2} \int d
\Omega \phi(x,y,z,\Omega) \exp(-i \Omega t)$, from Eq.(1) it
follows that the spectral amplitude $\phi(x,y,z,\Omega)$ satisfies
the wave equation
\begin{equation}
i \phi_z=- \frac{1}{2 \beta} (\phi_{xx}+\phi_{yy})+V(x,y)
\phi-\frac{\Omega}{v_g} \phi.
\end{equation}
Let us indicate by $u_1(\rho)$ and $u_2(\rho)$ the fundamental
modes of waveguides W$_1$ and W$_2$, and by $u_k(\rho)$ the
continuous set of modes of the slab waveguide S with the
normalization conditions $\int d\rho |u_1(\rho)|^2=\int d\rho
|u_2(\rho)|^2=1$, $\int d\rho
u_k(\rho)u_{k'}^*(\rho)=\delta(k-k')$ and $\int d\rho
u_1(\rho)u_2^*(\rho) \simeq \int d\rho u_1(\rho)u_k^*(\rho) \simeq
\int d\rho u_2(\rho)u_k^*(\rho) \simeq 0$, where $\rho \equiv
(x,y)$. Let us search for a solution to Eq.(3) in the form
\begin{eqnarray}
\phi(\rho,z,\Omega) & = &  \left[ c_1(z,\Omega)
u_1(\rho)+c_2(z,\Omega) u_2(\rho) + \right. \\
& + &  \left. \int dk c_k(z,\Omega) u_k(\rho)  \right] \exp(i
\Omega z / v_g). \nonumber
\end{eqnarray}
The evolution equations of modal amplitudes $c_1$, $c_2$ and $c_k$
then read \cite{Longhi08PRA}
\begin{eqnarray}
i \frac{ \partial c_1}{ \partial z} & = & \Delta \beta_0 c_1+\int dk g_1(k) c_k \\
i \frac{ \partial c_2}{ \partial z} & = & \Delta \beta_0 c_2+ \int dk g_2(k) c_k \\
i \frac{ \partial c_k}{ \partial z} & = & \Delta
\beta(k)c_k+g_1^*(k) c_1+g_2^*(k) c_2
\end{eqnarray}
where: $g_1(k)$ is the coupling amplitude between modes $u_1$ and
$u_k$; $g_2(k)$ is the coupling amplitude between modes $u_2$ and
$u_k$; and $\Delta \beta(k)$, $\Delta \beta_0$ are the propagation
constant shifts of modes $u_k$ and $u_{1,2}$, respectively, from
the reference value $\beta$. In their present form, Eqs.(5-7) are
analogous to the dynamical equations describing the quantum
mechanical decay of two bound states W$_1$ and W$_2$ into a common
continuum S [see Fig.1(b)], provided that the temporal dynamics of
the quantum mechanical problem is replaced by the paraxial
propagation in space of light waves. The two bound states, with
the same energy, are embedded in the continuum provided that
$\Delta \beta_0$ falls inside the continuous spectrum $\Delta
\beta(k)$, a condition which is satisfied whenever the refractive
index change $\Delta n_S$ in the slab waveguide is smaller than
the index change $\Delta n_g$ in the two channel waveguides (see
\cite{Longhi08PRA} for more details). Note that, if waveguides
W$_1$ and W$_2$ are symmetrically placed at opposite sides from
the slab waveguide S as shown in Fig.1(a), the coupling
coefficients $g_1(k)$ and $g_2(k)$ are the same, i.e.
$g_2(k)=g_1(k)\equiv g(k)$, and in this case a trapped state does
exist, as discussed in Ref.\cite{Longhi08PRA}. Although trapped
states in the continuum may also exist when the waveguides W$_1$
and W$_2$ are horizontally displaced or placed at different
distances from the slab waveguide \cite{Longhi08PRA}, in this work
we will consider the simplest symmetric case shown in Fig.1(a).

\subsection{Optical analogue of population trapping}

If one of the two channel waveguides, say e.g. W$_2$, were
removed, light initially injected into waveguide W$_1$ would decay
into the slab waveguide, a phenomenon fully analogous to the
quantum mechanical decay of a bound state coupled to a continuum.
In the markovian approximation, valid for weak coupling and for a
nearly unstructured continuum, the decay is well described by an
exponential law. The presence of waveguide W$_2$ generally
modifies the decay behavior and, under certain conditions, the
decay can be suppressed owing to a destructive Fano-like
interference of different decay channels. Full or fractional
suppression of the decay is related to the appearance of a trapped
(or dark) state in the continuum. To derive the decay laws of
light waves in the channel waveguides W$_1$ and W$_2$, we follow a
standard procedure \cite{Knight90}, detailed for instance in
Ref.\cite{Longhi08PRA}, and eliminate the amplitudes $c_k$ of
continuous modes by a formal integration of Eq.(7) with the
initial condition $c_k(0,\Omega)=0$. This yields a set of two
coupled integro-differential equations for the amplitudes $c_1$
and $c_2$ of discrete modes. In the weak coupling limit and
assuming a nearly unstructured continuum, the markovian
approximation can be made and the following reduced equations for
amplitudes $c_1$ and $c_2$ are derived
\begin{equation}
\frac{\partial c_1}{\partial z}=-i \Delta \beta_0
c_1-\sigma(c_1+c_2) \; , \; \; \frac{\partial c_2}{\partial z}=-i
\Delta \beta_0 c_2-\sigma(c_1+c_2),
\end{equation}
where
\begin{equation}
\sigma = \int_0^{\infty} d \tau \int dk |g(k)|^2 \exp \{-i[\Delta
\beta(k)-\Delta \beta_0] \tau \}
\end{equation}
is the decay rate of the single channel waveguide into the
continuum. The solution to Eqs.(8) reads explicitly
\begin{eqnarray}
c_1(z,\Omega) & = & S_{11}(z)c_1(0,\Omega)+S_{12}(z)c_2(0,\Omega) \\
c_2(z,\Omega) & = & S_{21}(z)c_1(0,\Omega)+S_{22}(z)c_2(0,\Omega)
\end{eqnarray}
where
\begin{eqnarray}
S_{11}(z) & = & \frac{1}{2} \exp(-i \Delta \beta_0 z)
 \left[1+\exp(-2 \sigma z)
\right]\\
S_{12}(z) & = & -\frac{1}{2} \exp(-i \Delta \beta_0 z)
 \left[1-\exp(-2 \sigma z) \right] \\
 S_{22}(z) & = & S_{11}(z) \; , \;\; S_{21}(z)=S_{12}(z).
\end{eqnarray}
Note that, in the quasi-monochromatic approximation assumed in
this work, the matrix coefficients $S_{n,l}$ are independent of
frequency $\Omega$. To understand the appearance of the optical
analogue of population trapping, let us consider the monochromatic
case, with the only nonvanishing spectral component at $\Omega=0$,
and two different input excitations, corresponding the former to
single waveguide excitation and the latter to simultaneous
excitation of the two channel waveguides. In the former case,
assuming for instance $c_1(0,\Omega)=\sqrt{2 \pi} \delta(\Omega)$
and $c_2(0,\Omega)=0$, one obtains
\begin{equation}
\psi(\rho,z)= S_{11}(z) u_1(\rho)+S_{12}(z) u_2(\rho)+S_{13}(z)
\theta (\rho,z)
\end{equation}
where the last term on the right hand side of Eq.(15) accounts for
the light field tunnelled into the slab waveguide and
$\theta(\rho,z)$ ia a suitable superposition of continuous modes
$u_k(\rho)$ normalized such that $\int d\rho
|\theta(\rho,z)|^2=1$. For power conservation, the relation
$|S_{11}|^2+|S_{12}|^2+|S_{13}|^2=1$ then holds. Note that, after
a propagation distance $z$ a few times the decay length $l_d
\equiv (1 / \sigma)$, one has $|S_{11}|^2=|S_{12}|^2=1/4$ and
$|S_{13}|^2 \simeq 1/2$, i.e. half of the injected light power has
decayed into the slab, whereas the other half of light power is
equally distributed into the two channel waveguides. The fact that
the decay is not complete (fractional decay) indicates that a
bound state embedded in the continuum does exist. When both
waveguides W$_1$ and W$_2$ are initially excited with coherent
fields of equal amplitudes but opposite sign, i.e.
 $c_1(0,\Omega)=-c_2(0,\Omega)= \sqrt{2 \pi} \delta(\Omega)$, one obtains $\psi(\rho,z)=\left[
u_1(\rho)-u_2(\rho) \right] \exp(-i \Delta \beta_0 z)$, i.e. the
decay into the slab waveguide is fully suppressed. This is due to
a destructive interference effect between different decay channels
when $c_2=-c_1$ [see Eqs.(7) and (8)] and to the existence of a
trapped state embedded in the continuum. Numerical examples of
fractional light decay for single waveguide excitation, and of
full decay suppression for simultaneous waveguide excitation in
the trapped state, as obtained by a direct numerical analysis of
Eq.(1) in the monochromatic regime, are shown in Fig.2. In the
simulations, we assumed circular channel waveguides with a
Gaussian index core profile of radius $r_c$ (at $1/e$), and a
step-index slab waveguide of thickness $a$. Equation (1) has been
integrated by a standard split-step pseudospectral
method with absorbing boundary conditions \cite{Longhi08PRA}.\\
 Generalization of light
propagation in the non-monochromatic case simply follows from the
superposition principle. For instance, if waveguides W$_1$ and
W$_2$ are excited at the input plane by two pulses with envelopes
$r_1(t)$ and $r_2(t)$, from Eqs.(4), (10) and (11) it follows that
the field envelope $\psi(\rho,z,t)$ can be cast in the form
\begin{widetext}
\begin{eqnarray}
\psi(\rho,z,t) = \left[ S_{11}(z)r_1 \left(t-\frac{z}{v_g}
\right)+S_{21}(z) r_2 \left(t-\frac{z}{v_g} \right)
\right]u_1(\rho)+ \nonumber \\
+\left[ S_{12}(z)r_1 \left(t-\frac{z}{v_g} \right)+S_{22}(z)r_2
\left(t-\frac{z}{v_g} \right) \right]u_2(\rho)+\psi_S
\end{eqnarray}
\end{widetext}
where $\psi_S=\psi_S(\rho,z,t)$ is the fractional part of the
field tunnelled into the slab waveguide. In particular, let us
consider two coherent pulses with the same envelope but phase
reversed and delayed by an interval $\delta$, i.e. $r_1(t)=r(t)$
and $r_2(t)=-r(t-\delta)$. In this case, interference effects,
leading to light trapping in waveguides W$_1$ and W$_2$ and full
suppression of leakage into the slab waveguide S ($\psi_S \simeq
0$), occurs for $\delta=0$ or, approximately, for a delay $\delta$
much smaller than the characteristic pulse duration $\tau_c$. For
$\delta$ larger than $\tau_c$, the two wave packets are not
overlapped and behave as independent beams, leading to fractional
decay. The latter scenario is also observed if the two pulses are
temporally overlapped $(\delta=0)$ but incoherent, i.e. their
phase difference changes randomly in time.

\section{Quantum description and nonclassical effects}
\subsection{Quantization procedure}
To describe propagation of nonclassical light in the coupled
waveguide system, the classical paraxial wave field $\psi$ in
Eq.(1) or, similarly, the classical c-numbers $c_{1}$, $c_2$ and
$c_k$ in Eqs.(5-7), have to be replaced by quantum-mechanical
operators satisfying suitable commutation relations, and different
quantization procedures may be adopted. A first approach, which is
well suited when the classical problem is formulated in terms of
coupled-mode equations (5-7), is the input-output operator
formalism commonly used for linear quantum-optical networks,
either in the Heisenberg or in the Schr\"{o}dinger pictures (see,
for instance, \cite{Lai91,Leonhard03}). A second approach, suited
when the classical problem in Eq.(1) is formulated as a
propagative (rather than as an initial-value) problem, is to adopt
a quantization procedure for the classical field $\psi$ as an
evolution in space (rather than in time). In the Schr\"{o}dinger
picture, this leads to an evolution in space of a many-photon
probability amplitude. Such a phenomenological approach, which
will be adopted in the following analysis, has received a growing
use and appreciation in quantum theories of optical solitons
\cite{Lai89,Matsko00,varialtri}; its consistency with standard
canonical quantization procedure has been discussed in
Ref.\cite{Matsko00}. Similar procedures have been also developed
to study, in the Heisenberg picture, paraxial propagation of
nonclassical light and applied to problems of quantum imaging
(see, for instance, \cite{imaging}). The quantization procedure
consists in writing the classical paraxial wave equation (1) in
Hamiltonian form assuming the paraxial spatial coordinate $z$ as
an independent variable \cite{Lai89}. Introducing the new field
$\Pi=i \hbar \psi^\dag$ and the Hamiltonian $H=\int dx dy dt
\mathcal{H}$ with density
\begin{equation}
\mathcal{H}=-\frac{i}{2 \beta} (\Pi_x \psi_x+\Pi_y
\psi_y)-\frac{1}{v_g} \Pi \psi_t-iV(x,y)\Pi \psi,
\end{equation}
 it readily follows that the Hamilton
equations $ \psi_z= (\delta H/ \delta \Pi)$, $\Pi_z= -(\delta H/
\delta \psi)$ yield Eq.(1) and its complex conjugate, so that
$\Pi$ is canonically conjugated to $\psi$. Quantization is then
accomplished by replacing the classical fields $\psi$ and $\Pi$
with the operators $\hat{\psi}(x,y,t)$ and $\hat{\Pi}= i \hbar
\hat{\psi ^\dag}(x,y,t)$ satisfying the commutation relations
$[\hat{\psi}(\rho,t),\hat{\psi^\dag}(\rho',t')]=\delta(\rho-\rho')
\delta(t-t')$ and
$[\hat{\psi}(\rho,t),\hat{\psi}(\rho',t')]=[\hat{\psi^{\dag}}(\rho,t),\hat{\psi^\dag}(\rho',t')]=0$,
where  we have set $\rho=(x,y)$ . By introducing the spectral
decomposition $\hat{\psi}(\rho,t)=(2 \pi)^{-1/2} \int d \Omega
\hat{\phi}(\rho, \Omega) \exp(-i\Omega t)$, the second-quantized
Hamiltonian operator reads
 \begin{equation}
 \hat{H}= \hbar \int d \rho d \Omega \left\{ \frac{1}{2 \beta} (\hat{\phi}^{\dag}_{x} \hat{\phi}_x+\hat{\phi}^{\dag}_{y} \hat{\phi}_y) +
 \left[V(\rho)-\frac{\Omega}{v_g} \right] \hat{\phi}^\dag \hat{\phi}
 \right\}.
 \end{equation}
Note that in the spectral domain the following commutation
relations hold for the operators $\hat \phi(\rho,\Omega)$ and
$\hat {\phi }^\dag(\rho,\Omega)$
\begin{eqnarray}
\left[
\hat{\phi}(\rho,\Omega),\hat{\phi^\dag}(\rho',\Omega')\right] =
\delta(\rho-\rho') \delta(\Omega-\Omega') , \nonumber \\
 \left[ \hat{\phi}(\rho,\Omega), \hat{\phi}(\rho',\Omega') \right]   =  \left[
\hat{\phi^\dag }(\rho,\Omega),\hat{\phi ^\dag}(\rho',\Omega')
\right]=0.
\end{eqnarray}
 Note also that
the field energy $U$ [Eq.(2)] corresponds to the operator
$\hat{U}= \hbar \omega \int d \rho d \Omega
\hat{\phi}^\dag(\rho,\Omega) \hat{\phi}(\rho,\Omega)$. In the
Schr\"{o}dinger picture, the quantum field is described by a
vector state $|\mathcal{Q}(z)\rangle$ which evolves according to
\begin{equation}
i \hbar \frac{d |\mathcal{Q}\rangle}{dz}= \hat{H}
|\mathcal{Q}\rangle
\end{equation}
whereas the operators $\hat{\phi}(\rho,\Omega)$ do not evolve with
$z$ \cite{note1}. The state $|\mathcal{Q}\rangle$ can be expanded
in Fock space as $|\mathcal{Q}\rangle=\sum_n a_n |f^{(n)}({\mathbf
q},\mathbf{\Omega},z)\rangle$, where the $n$-photon number state
$|f^{(n)} \rangle$ is defined by (see, for instance, \cite{Lai89})
\begin{equation}
|f^{(n)} \rangle  =  \int d{\mathbf q} d \mathbf{\Omega}
\frac{f^{(n)}(\mathbf{q}, \mathbf{\Omega},z)}{\sqrt{n!}}
  \hat{\phi^\dag}(\rho_1,\Omega_1)....
\hat{\phi^\dag}(\rho_n,\Omega_n) |0\rangle
\end{equation}
and where we have set $\mathbf{q}=(\rho_1,\rho_2,...)$,
$\mathbf{\Omega}=(\Omega_1,\Omega_2,...)$. The normalization
conditions $\sum_n|a_n|^2=1$ and $\int d{\mathbf q} d
\mathbf{\Omega} |f^{(n)}(\mathbf{q},\mathbf{\Omega},z)|^2=1$ are
also assumed, which ensure that $\langle
\mathcal{Q}|\mathcal{Q}\rangle=1$. Note that $\hat{U}|f^{(n)}
\rangle=n \hbar \omega |f^{(n)}\rangle$, i.e. the Fock state
$|f^{(n)} \rangle$ is obtained from the vacuum state $|0\rangle$
by creating $n$ photons with space-frequency weighting function
$f^{(n)}$. The evolution equation for the weighting function
$f^{(n)}$ is obtained by substituting Eqs.(18) and (21) into the
Scr\"{o}dinger equation (20) and using the commutation relations
of field operators [Eqs.(19)]. One then obtains
\begin{equation}
i \frac{\partial f^{(n)}}{\partial z}=\sum_{l=1}^{n} \left[
-\frac{1} {2 \beta}
 \left( \frac {
\partial^2}{\partial x_l^2}+\frac {
\partial^2}{\partial y_l^2} \right) +V(\rho_l)-\frac{\Omega_l}{v_g} \right] f^{(n)}.
\end{equation}
Owing to the bosonic nature of photons, solely symmetric functions
$f^{(n)}$ should be considered.

\subsection{Nonclassical effects with monochromatic beams}

 Let us consider first the propagation of monochromatic fields, so that in
Eqs.(18), (19), (21) and (22) we may disregard integration over
different spectral components $\Omega$ and use a single
renormalized bosonic creation operator $\hat{\phi}^\dag(\rho)$ at
frequency $\Omega=0$ satisfying the commutation relations
\cite{note2}
 $[\hat{\phi}(\rho),\hat{\phi}^\dag(\rho')]=\delta(\rho-\rho')$ and $[\hat{\phi}(\rho),\hat{\phi}(\rho')]=
  [\hat{\phi}^\dag (\rho),\hat{\phi}^\dag(\rho')]=0$ , which
  replace Eq.(19).
The simplest $n$-photon number state, denoted by $|g\rangle_n$, is
obtained by assuming in Eq.(22)
$f^{(n)}(\mathbf{q},z)=g(\rho_1,z)g(\rho_2,z)... g(\rho_n,z)$,
where the function $g(\rho,z)$ satisfies the classical wave optics
equation
\begin{equation}
i g_z=- \frac{1}{2 \beta} (g_{xx}+g_{yy})+V(x,y) g
\end{equation}
with the normalization $\int dx dy |g(x,y,z)|^2=1$. In this case
one has
\begin{equation}
|g\rangle_n=\frac{1}{\sqrt {n!}} \left( \int dx dy g(x,y,z)
\hat{\phi^\dag}(x,y) \right)^n |0 \rangle.
\end{equation}
Physically, this state describes the excitation of the optical
system with an $n$-photon number state input beam with a spatial
profile $g(x,y,0)$ at the entrance plane $z=0$. The classic wave
optics description of light propagation, discussed in Sec.II, is
attained by considering a superposition of photon number states
$|g_n\rangle$ with a Poissonian distribution with c-number
$\alpha$, i.e. the coherent state $|g; \alpha \rangle_{{\rm coh}}$
defined by
\begin{equation}
|g;\alpha \rangle_{{\rm coh}}=\sum_{n=0}^{\infty}
\frac{\exp(-|\alpha|^2/2) \alpha^n}{n!} \left( \int d \rho
g(\rho,z) \hat \phi^{\dag}(\rho) \right)^n |0\rangle.
\end{equation}
One can readily show that the coherent state $|g; \alpha
\rangle_{{\rm coh}}$ is an eigenstate of the field annihilation
operator $\hat{\phi}(x,y)$ with eigenvalue $\alpha g(x,y,z)$, i.e.
$\hat{\phi}(x,y) |g; \alpha \rangle_{{\rm coh}}=\alpha g(x,y,z)
|g; \alpha \rangle_{{\rm coh}}$, where $g(x,y,z)$ evolves
according to the classical wave equation (23). Therefore the
expectation value $_{\rm coh}\langle g; \alpha| \hat{\phi}(x,y)
|g; \alpha \rangle_{{\rm coh}}=\alpha g(x,y,z)$ yields the
classical solution of the wave equation (23) for an input beam
profile $\alpha g(x,y,0)$. More generally,  for a nonclassical
state $|\mathcal{Q} \rangle$ obtained by an arbitrary
superposition of photon number states $|g\rangle_n$ with
amplitudes $a_n$, one can readily show that the expectation value
of $\hat{\phi}^\dag (x,y) \hat{\phi}(x,y)$ yields the classic wave
optics intensity distribution, namely $\langle
\mathcal{Q}|\hat{\phi}^\dag (x,y) \hat{\phi}(x,y) | \mathcal{Q}
\rangle=\langle n \rangle |g(x,y,z)|^2$, where $\langle n
\rangle=\sum_n n |a_n|^2$ is the mean photon number of the input
beam. The quantum aspects of nonclassical light for single beam
excitation may be revealed when the statistics of photons trapped
in waveguides  W$_1$ and  W$_2$ are considered. For example, let
us consider excitation of waveguide  W$_1$ in its fundamental mode
at the input plane $z=0$, so that $g(x,y,0)=u_1(x,y)$, and let us
compare the statistics of photons that remain in waveguide W$_1$
when the input beam is a photon number state (nonclassical light)
or a coherent state (classical light).  According to the analysis
of Sec.II.B, the wave amplitude $g$ evolves according to [see
Eq.(15)]
\begin{equation}
g(\rho,z)=S_{11}(z)u_1(\rho)+S_{12}u_2(\rho)+S_{13}(z)\theta(\rho,z),
\end{equation}
where $|S_{11}|^2+|S_{12}|^2+|S_{13}|^2=1 $ and $\theta(\rho,z)$
defines the normalized spatial profile of the field tunnelled into
the slab waveguide. If we introduce the operators
\begin{eqnarray}
\hat {a}^{\dag}_{1} & \equiv & \int dx dy u_{1}(x,y) \hat \phi^{\dag}(x,y) \\
\hat {a}^{\dag}_{2} & \equiv & \int dx dy u_{2}(x,y) \hat \phi^{\dag}(x,y) \\
\hat {a}^{\dag}_{3} & \equiv & \int dx dy  \theta(x,y,z) \hat
\phi^{\dag}(x,y)
\end{eqnarray}
the commutation relations $[\hat {a}_{i},\hat
{a}_{k}^{\dag}]=\delta_{i,k}$ and $[\hat {a}_{i}^{\dag},\hat
{a}_{k}^{\dag}]=[\hat {a}_{i},\hat {a}_{k}]=0$ ($i,k=1,2,3$) hold.
Assuming that the vector state $|\mathcal{Q} \rangle$ is given by
a superposition of photon number states $|g\rangle_n$ with
amplitudes $a_n$, one can write
\begin{equation}
|\mathcal{Q} \rangle=\sum_{n=0}^{\infty} \frac{a_n}{\sqrt{n!}}
\left(S_{11} a_1^{\dag}+S_{12} a_2^{\dag}+S_{13} a_3^{\dag}
\right)^n |0\rangle.
\end{equation}
The joint photon distribution $P(n_1;n_2,n_3;z)$ to find $n_1$
photons in waveguide W$_1$, $n_2$ photons in waveguide W$_2$ and
$n_3$ photons in the slab waveguide S is given by
\begin{equation}
P(n_1,n_2,n_3;z)=|\langle n_1,n_2,n_3| \mathcal{Q} \rangle|^2
\end{equation}
where we have set
\begin{equation}
|n_1,n_2,n_3 \rangle \equiv \frac{1}{\sqrt{n_1! n_2! n_3!}} \hat
a_{1}^{\dag n_1} \hat a_{2}^{\dag n_2} \hat a_{3}^{\dag n_3} |0
\rangle.
\end{equation}
The explicit expression of $P(n_1,n_2,n_3;z)$ can be obtained by a
double binomial expansion of the operator $(S_{11}
a_1^{\dag}+S_{12} a_2^{\dag}+S_{13} a_3^{\dag})^n$ entering in
Eq.(30). The marginal photon distribution $P_1(n_1,z)$ to find
$n_1$ photons in waveguide W$_1$ is then obtained as
$P_1(n_1;z)=\sum_{n_2,n_3=0}^{\infty} P(n_1,n_2,n_3;z)$, and reads
explicitly (see, for instance, \cite{Lai91})
\begin{equation}
P_1(n_1;z)=\sum_{n=n_1}^{\infty} |a_n|^2 \left(
\begin{array}{c}
n \\
n_1
\end{array}
\right) \kappa^{n_1} (1-\kappa)^{n-n_1}
\end{equation}
where $\kappa \equiv |S_{11}(z)|^2$. Similar expressions are
obtained for the marginal photon distributions $P_2(n_2,z)$ and
$P_3(n_3,z)$ for waveguides W$_2$ and $S$ after setting
$\kappa=|S_{12}(z)|^2$ and $\kappa=|S_{13}(z)|^2$, respectively.
Let us now suppose that the input beam is a coherent state with
mean photon number $\langle n \rangle=n_0$ and Poissonian
distribution
\begin{equation}
|a_n|^2=\frac{n_0^n  \exp(- n_0)}{n!}.
\end{equation}
From Eq.(33) it follows that the marginal photon distribution
$P_1(n_1;z)$ remains Poissonian with mean photon number $\langle
n_1 \rangle=n_0 |S_{11}(z)|^2$ that decays along the propagation
distance $z$ according to the classical decay law given by
Eq.(12). Conversely, if the input beam is a photon number state,
i.e. $a_{n}=\delta_{n,n_0}$, from Eq.(33) it follows that the
marginal photon distribution $P_1(n_1;z)$ is given by the binomial
distribution
\begin{equation}
P_1(n_1;z)=\left(
\begin{array}{c}
n_0 \\
n_1
\end{array}
\right) \kappa^{n_1} (1-\kappa)^{n_0-n_1} \; \; (n_1 \leq n_0)
\end{equation}
[$P_1(n_1;z)=0$ for $n_1 >n_0$] with photon mean $\langle n_1
\rangle=n_0|S_{11}(z)|^2$ and variance $\langle \Delta n_{1}^2
\rangle=|S_{11}(z)|^2[1-|S_{11}(z)|^2]n_0$. The binomial photon
distribution highlights the particle-like behavior of photons
undergoing the decay from waveguide W$_1$ and it is analogous to
that created by a beam splitter when excited by a photon number
state in one port, and the vacuum state in the other one (see, for
instance, \cite{Campos89}). The photons in the initially excited
waveguide behave like independent classical particles and tunnel
into the other waveguides is ruled by a Bernoulli trial (a coin
toss) with a cumulative tunneling probability  given by $1-|S_{11}(z)|^2$.\\
When the two waveguides W$_1$ and W$_2$ are simultaneously excited
by two nonclassical independent beams, quantum signatures of light
propagation resulting from quantum interference can be detected by
photon coincidence measurements. To describe propagation of
independent beams, let us notice that for a given set of
normalized and orthogonal solutions $g_1(\rho,z)$, $g_2(\rho,z)$,
... to the classic wave equation (23), one can construct the
$n$-photon number state
\begin{widetext}
\begin{equation}
| g_1,g_2,...\rangle_{n_1,n_2,...}= \frac{1}{\sqrt{n_1! n_2!
....}} \left(\int d \rho g_1(\rho,z) \hat \phi^{\dag}(\rho)
\right)^{n_1} \times \left(\int d \rho g_2(\rho,z) \hat
\phi^{\dag}(\rho) \right)^{n_2} \times ... |0\rangle
\end{equation}
\end{widetext}
 which describes excitation of the optical system with a set of independent
 beams with spatial profiles $g_1$, $g_2$,... carrying $n_1$,
$n_2$,... photons ($n=n_1+n_2+...$). In particular, let us assume
that at the input plane the waveguides W$_1$ and W$_2$ are excited
in their fundamental modes by single photon number states, i.e.
$g_1(\rho,0)=u_1(\rho)$, $g_2(\rho,0)=u_2(\rho)$ and
\begin{equation}
|Q(z=0)\rangle=|u_1,u_2\rangle_{1,1}=\hat {a}^{\dag}_1 \hat
{a}^{\dag}_2 |0\rangle,
\end{equation}
where the operators $\hat a^{\dag}_1$ and $\hat a^{\dag}_2$ are
defined by Eqs.(27) and (28). According to the classical analysis
of Sec.II.B, the waves $g_1$ and $g_2$ evolve according to
\begin{eqnarray}
g_1(\rho,z) & = &
S_{11}(z)u_1(\rho)+S_{12}(z)u_2(\rho)+S_{13}(z)\theta(\rho,z)
\nonumber \\
& & \\
 g_2(\rho,z) & = &
S_{21}(z)u_1(\rho)+S_{22}(z)u_2(\rho)+S_{23}(z)\theta(\rho,z)
\nonumber \\
& &
\end{eqnarray}
with $S_{23}=S_{13}$ for symmetry reasons \cite{note3},
$S_{12}=S_{21}$, $S_{11}=S_{22}$ [see Eqs.(12-14)], and
$|S_{11}|^2+|S_{12}|^2+|S_{13}|^2=1$ for power conservation. From
Eqs.(36), (38) and (39) it follows that the state vector of the
system at the generic propagation distance $z$ is given by
\begin{equation}
|\mathcal{Q} \rangle=\left (S_{11} \hat{a}^{\dag}_{1} +S_{12}
\hat{a}^{\dag}_{2} + S_{13} \hat{a}^{\dag}_{3} \right) \left(
S_{21} \hat{a}^{\dag}_{1} +S_{22} \hat{a}^{\dag}_{2} + S_{23}
\hat{a}^{\dag}_{3} \right) |0\rangle
\end{equation}
where $\hat{a}^{\dag}_3$ is defined in Eq.(29). It can be readily
shown that the expectation value of the field intensity for the
two-photon state (40) is given by the incoherent superposition
\begin{equation}
\langle \mathcal{Q}|\hat{\phi}^{\dag}(\rho) \hat{\phi}(\rho)
|\mathcal{Q}\rangle=|g_1(\rho,z)|^2+|g_2(\rho,z)|^2,
\end{equation}
 the absence of interference being
due to the lack of a definite phase relationship between the two
photons (see, for instance, \cite{Mandel87}). The two independent
beams which excite the two waveguides W$_1$ and W$_2$ behave,
therefore, as two incoherent classical fields, and therefore
according to the analysis of Sec.II.B half of the light power
tunnels, on average, into the slab waveguide S for propagation
distances $z \gg l_d$ . As the operators $\hat{a}^{\dag}_{1}$,
$\hat{a}^{\dag}_{2}$ and $\hat{a}^{\dag}_{3}$ create photons in
waveguides W$_1$, W$_2$ and W$_3$, respectively [see Eqs.(27-29)],
from Eq.(40) the joint photon distribution $P(n_1,n_2,n_3;z)$ can
be readily calculated, and its nonvanishing terms are explicitly
given by
\begin{widetext}
\begin{eqnarray}
P(2,0,0;z) & = &  P(0,2,0;z) = 2 |S_{11}|^2 |S_{21}|^2= \frac{1}{8} \left[1-\exp(-4 \sigma z) \right]^2 \\
P(1,1,0;z) & = & |S_{11}S_{22}+S_{12}S_{21}|^2= \frac{1}{4} \left[1+\exp(-4 \sigma z) \right]^2  \\
P(1,0,1;z) & = & P(0,1,1;z)=|S_{11}S_{23}+S_{13}S_{21}|^2= \frac{1}{2} \exp(-4 \sigma z) \left[1-\exp(-4 \sigma z) \right]  \\
P(0,0,2;z)& = & 2 |S_{13}|^2|S_{23}|^2= \frac{1}{2}
\left[1-\exp(-4 \sigma z) \right]^2.
\end{eqnarray}
\end{widetext}
An important result is that, for propagation distances $z$ larger
than the characteristic decay length $l_d= (1 / \sigma)$, the
joint probability $P(1,0,1)$ to find one photon in waveguide W$_1$
and one in the slab waveguide S vanishes, and similarly for
$P(0,1,1)$ [see Eq.(44)]. Such a result indicates that, if one of
the two initially injected photons tunnels into the slab
waveguide, the other photon does the same, i.e. photon pairs bunch
when decaying into the continuum. This is obviously a nonclassical
effect similar to the two-photon Hong-Ou-Mandel quantum
interference in a 50$\%$ beam splitter \cite{Hong87}. In our case,
the vanishing of $P(1,0,1)$ [and similarly of $P(0,1,1)$] is
related to a destructive interference between the probability
amplitudes $S_{11}S_{23}$ and $S_{13}S_{21}$ entering in Eq.(44)
which describe two possible paths for the photon pair.\\

\subsection{Nonclassical effects with wave packets}

In the previous subsection, we have considered propagation of
nonclassical light in the monochromatic limit, however in practice
nonclassical light such as that generated by spontaneous
parametric down conversion in a nonlinear crystal is always
polychromatic to some extent. Moreover, in experimental settings
quantum interference such as the two-photon Hong-Ou-Mandel
interference is typically revealed as a dip in intensity
correlation (photon coincidence) measurements when the time delay
between the two incoming wave packets is varied
\cite{Hong87,Politi08}. It is therefore useful to extend the
analysis of Sec.III.B to the
case of polychromatic wave packets.\\
For the sake of definiteness, let us assume that waveguides W$_1$
and W$_2$ are excited in their fundamental spatial modes by two
spectrally-narrow wave packets generated by spontaneous parametric
down conversion in type-I nonlinear crystal. In this case, the
vector state $|\mathcal{Q} \rangle$ at the input plane of the
waveguiding system is a generalization of Eq.(37) and given by a
superposition of two-photon states with a spectral function
$C(\Omega_1,\Omega_2)$ (see, for instance, \cite{Hong87,Wang06})
\begin{equation}
|\mathcal{Q} (z=0) \rangle = \int d \Omega_1 d \Omega_2
C(\Omega_1, \Omega_2) \hat{a}_{1}^{\dag}(\Omega_1)
\hat{a}_{2}^{\dag}(\Omega_2) |0\rangle
\end{equation}
where the operators $\hat{a}_{1}^{\dag}(\Omega_1)$ and
$\hat{a}_{2}^{\dag}(\Omega_1)$ are defined by
\begin{equation}
\hat{a}_{1,2}^{\dag}(\Omega) = \int d\rho \; u_{1,2}(\rho)
\hat{\phi}^{\dag}(\rho,\Omega).
\end{equation}
The evolution of vector state along the propagation distance $z$
can be obtained by a straightforward extension of the analysis of
Sec.III.B and reads explicitly
\begin{widetext}
\begin{eqnarray}
|\mathcal{Q} \rangle & = &  \int d \Omega_1 d \Omega_2 C(\Omega_1,
\Omega_2) \exp \left[i(\Omega_1 +\Omega_2) z/v_g \right] \left[
S_{11}(z) \hat{a}_{1}^{\dag}(\Omega_1)+ S_{12}(z)
\hat{a}_{2}^{\dag}(\Omega_1) +S_{13}(z)
\hat{a}_{3}^{\dag}(\Omega_1) \right] \times \nonumber \\
& \times & \left[ S_{21}(z) \hat{a}_{1}^{\dag}(\Omega_2)+
S_{22}(z) \hat{a}_{2}^{\dag}(\Omega_2) +S_{23}(z)
\hat{a}_{3}^{\dag}(\Omega_2) \right]
 |0\rangle
\end{eqnarray}
\end{widetext}
where
\begin{equation}
\hat{a}_{3}^{\dag}(\Omega) = \int d\rho \; \theta(\rho,z)
\hat{\phi}^{\dag}(\rho,\Omega).
\end{equation}
To study quantum interference effects with wave packets, and in
particular the vanishing of the joint photon probability
$P(1,0,1;z)=P(0,1,1;z)$ for $z \gg l_d$ found for monochromatic
beams [Eq.(44)], let us introduce the integrated intensity
correlation function at times $t_1$ and $t_2$ in waveguides W$_1$
and S defined as
\begin{widetext}
\begin{equation}
\mathcal{P}(t_1,t_2;z)=\int_{\mathcal{A}_1} d \rho_1
\int_{\mathcal{A}_2} d \rho_2 \langle
\mathcal{Q}|\hat{\psi}^{\dag}(\rho_2,t_2)\hat{\psi}^{\dag}(\rho_1,t_1)
\hat{\psi}(\rho_1,t_1) \hat{\psi}(\rho_2,t_2)| \mathcal{Q}\rangle,
\end{equation}
\end{widetext}
where $\mathcal{A}_1$ and  $\mathcal{A}_2$ are two areas in the
transverse $(x,y)$ plane surrounding waveguides W$_1$ and S,
respectively. The correlation function $\mathcal{P}(t_1,t_2)$ is
proportional to the joint probability of detecting one photon in
waveguide W$_1$ at time $t_1$, and one photon in waveguide S at
time $t_2$, after a propagation distance $z$ from the input plane.
In fact, $\langle
\mathcal{Q}|\hat{\psi}^{\dag}(\rho_2,t_2)\hat{\psi}^{\dag}(\rho_1,t_1)
\hat{\psi}(\rho_1,t_1) \hat{\psi}(\rho_2,t_2)| \mathcal{Q}\rangle$
is proportional to the joint probability of detecting one photon
at point $\rho_1$ and time $t_1$, and one photon at point $\rho_2$
and time $t_2$ at the same propagation distance $z$. The integral
over the areas $\mathcal{A}_1$ and  $\mathcal{A}_2$ thus gives the
joint probability of detecting one photon in waveguide W$_1$ at
time $t_1$, and one photon in waveguide S at time $t_2$. For
photons generated by spontaneous parametric down conversion in
type-I nonlinear crystal using a monochromatic pump beam at
frequency $\omega_p=2 \omega$, the two-photon spectrum
$C(\Omega_1,\Omega_2)$ of a pair of signal and idler photons can
be expressed as \cite{Hong87,Wang06}
\begin{equation}
C(\Omega_1,\Omega_2)=\delta(\Omega_1+\Omega_2)G(\Omega_1,\Omega_2)
\exp \left[ i \frac{\delta(\Omega_1 -\Omega_2)}{2} \right]
\end{equation}
where $G(\Omega_1,\Omega_2)$ is the phase matching function. The
last exponential term on the right hand side in Eq.(51) has been
introduced to account for a possible time delay $\delta$ between
signal and idler wave packets introduced by different optical
paths from the crystal to waveguides W$_1$ and W$_2$. The phase
matching function $G(\Omega_1,\Omega_2)$ is assumed to be a
real-valued and symmetric function [i.e.
$G(\Omega_1,\Omega_2)=G(\Omega_2,\Omega_1)$], peaked at around
$\Omega_1=\Omega_2=0$, with e.g. a Gaussian profile \cite{Hong87}.
Taking into account that $\hat{\psi}(\rho,t)=(2 \pi)^{-1/2} \int d
\Omega \hat {\phi}(\rho, \Omega) \exp(-i \Omega t)$, substitution
of Eqs.(48) and (51) into Eq.(50), using the commutaion relations
$[\hat {\phi}(\rho,
\Omega),\hat{a}^{\dag}_{1,2}(\Omega')]=u_{1,2}(\rho)
\delta(\Omega-\Omega')$, $[\hat {\phi}(\rho,
\Omega),\hat{a}^{\dag}_{3}(\Omega')]=\theta(\rho,z)
\delta(\Omega-\Omega')$ and the relations $\int_{\mathcal{A}_{1}}
d \rho |u_{1}(\rho)|^2 \simeq 1$, $\int_{\mathcal{A}_2} d \rho
|\theta(\rho,z)|^2 \simeq 1$, $\int_{\mathcal{A}_{1,2}} d \rho
|u_{2,1}(\rho)|^2=\int_{\mathcal{A}_1} d \rho |\theta(\rho,z)|^2=
\int_{\mathcal{A}_2} d \rho |u_2(\rho)|^2 \simeq 0$, after some
lengthy but straightforward calculations one obtains
\begin{widetext}
\begin{equation}
\mathcal{P}(t_1,t_2;z)= \left| r(\tau+\delta) S_{11}(z) S_{23}(z)
+r(\tau-\delta) S_{13}(z)S_{21}(z) \right|^2
\end{equation}
\end{widetext}
where $\tau=t_2-t_1$ and where we introduced the real-valued
correlation function $r(\tau)$ defined by
\begin{equation}
r(\tau)=\frac{1}{2 \pi} \int d \Omega G(\Omega,-\Omega) \exp(-i
\Omega \tau).
\end{equation}
In practice, coincidence measurements correspond to an integration
of $\mathcal{P}(t_1,t_2;z)$ with respect to the time difference
$\tau=t_2-t_1$ over the resolving coincidence time, which is
typically much longer than the correlation time $\tau_c$ of
$g(\tau)$. Integrating Eq.(52) with respect to $\tau$ from
$-\infty$ to $\infty$ and taking into account that $S_{21}=S_{12}$
and $|S_{23}|^2=|S_{13}|^2=1-|S_{11}|^2-|S_{12}|^2$, the following
expression for the correlation function $\mathcal{P}$ versus time
delay $\delta$ is finally obtained
\begin{widetext}
\begin{eqnarray}
\mathcal{P}(\delta;z)=\alpha \left[1- |S_{11}(z)|^2-|S_{12}(z)|^2
\right] \left[ |S_{11}(z)|^2+ |S_{21}(z)|^{2} + \right. \nonumber
\\ + \left. 2 {\rm Re} \left[ S_{11}(z)S_{21}^*(z) \right]
\frac{\int_{-\infty}^{\infty} d \tau
r(\tau-\delta)r(\tau+\delta)}{\int_{-\infty}^{\infty} d \tau
r^2(\tau) }\right]
\end{eqnarray}
\end{widetext}
where $\alpha=\int_{-\infty}^{\infty} d \tau r^2(\tau) $ and the
expression of the coefficients $S_{11}(z)$ and $S_{12}(z)$ are
given by Eqs.(12) and (13). The behavior of
$\mathcal{P}(\delta;z)$ versus $\delta$ shows a characteristic dip
at $\delta=0$ of width $\sim \tau_c$, which is analogous to the
Hong-Ou-Mandel dip observed in two-photon interference from a beam
splitter \cite{Hong87}. Far from the dip, $\mathcal{P}(\delta)$
reaches a constant value
$\alpha(1-|S_{11}|^2-|S_{12}|^2)(|S_{11}|^2+|S_{21}|^2)$. For $z$
larger than the characteristic decay length $l_d=1/\sigma$, the
correlation $\mathcal{P}(\delta;z)$ vanishes at the dip
$\delta=0$, i.e. when signal and idler wave packets are temporally
overlapped, according to the analysis of Sec.III.B.

\section{Conclusions}
In this work we have theoretically investigated propagation of
classical and nonclassical light
 in a waveguide-based photonic
structure that provides an optical analogue of population trapping
in the continuum encountered in atomic physics. For classical
light waves, coupled-mode equation analysis, previously studied in
Ref.\cite{Longhi08PRA}, shows that Fano interference between
different light leakage channels is responsible for the appearance
of a trapped state embedded in the continuum. To study propagation
of nonclassical light, a second quantization model for the scalar
wave equation, in the paraxial and quasi-monochromatic
approximations, has been adopted. As for input beam excitation in
a coherent state the classical picture of light propagation is
retrieved, quantum interference effects with no classical
counterpart have been highlighted for photon number state
excitation of the waveguide structure. In particular, the tendency
of photon pairs to bunch when decaying into the continuum has been
predicted. Such an effect, which is similar to the two-photon
Hong-Ou-Mandel interference in a beam splitter \cite{Hong87}, may
be observed as a dip in photon coincidence measurements. Our
results indicate that photonic structures originally designed to
mimic with optical waves the classical analogues of
quantum-mechanical phenomena encountered in atomic, molecular or
condensed-matter physics \cite{Longhi09}, may exhibit themselves a
strictly quantum behavior when single photon level is reached.

\begin{figure}
\includegraphics[scale=0.4]{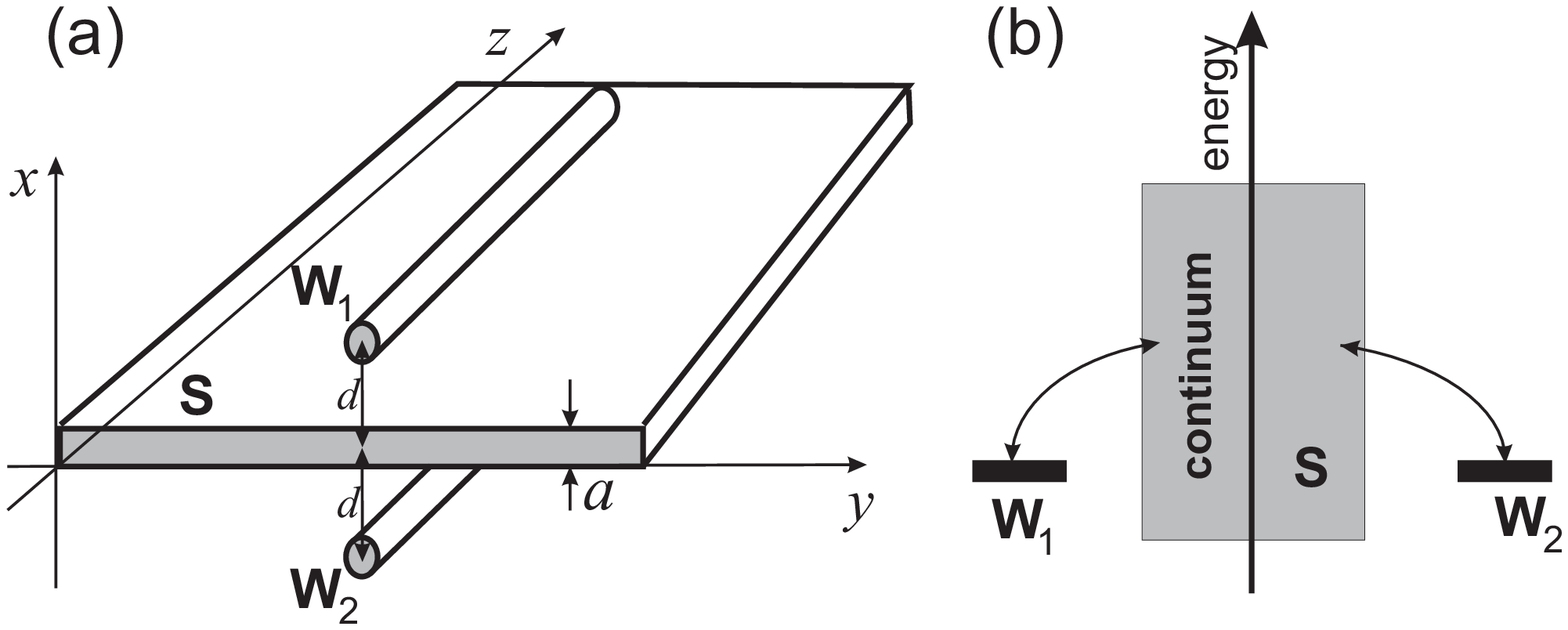}
\caption{(a) Schematic of the coupled waveguide system composed by
two equal single-mode channel waveguides W$_1$ and W$_2$
side-coupled to a slab waveguide S. (b) The quantum mechanical
analogue, showing the decay of two bound states coupled to a
common continuum.}
\end{figure}
\begin{figure}
\includegraphics[scale=0.6]{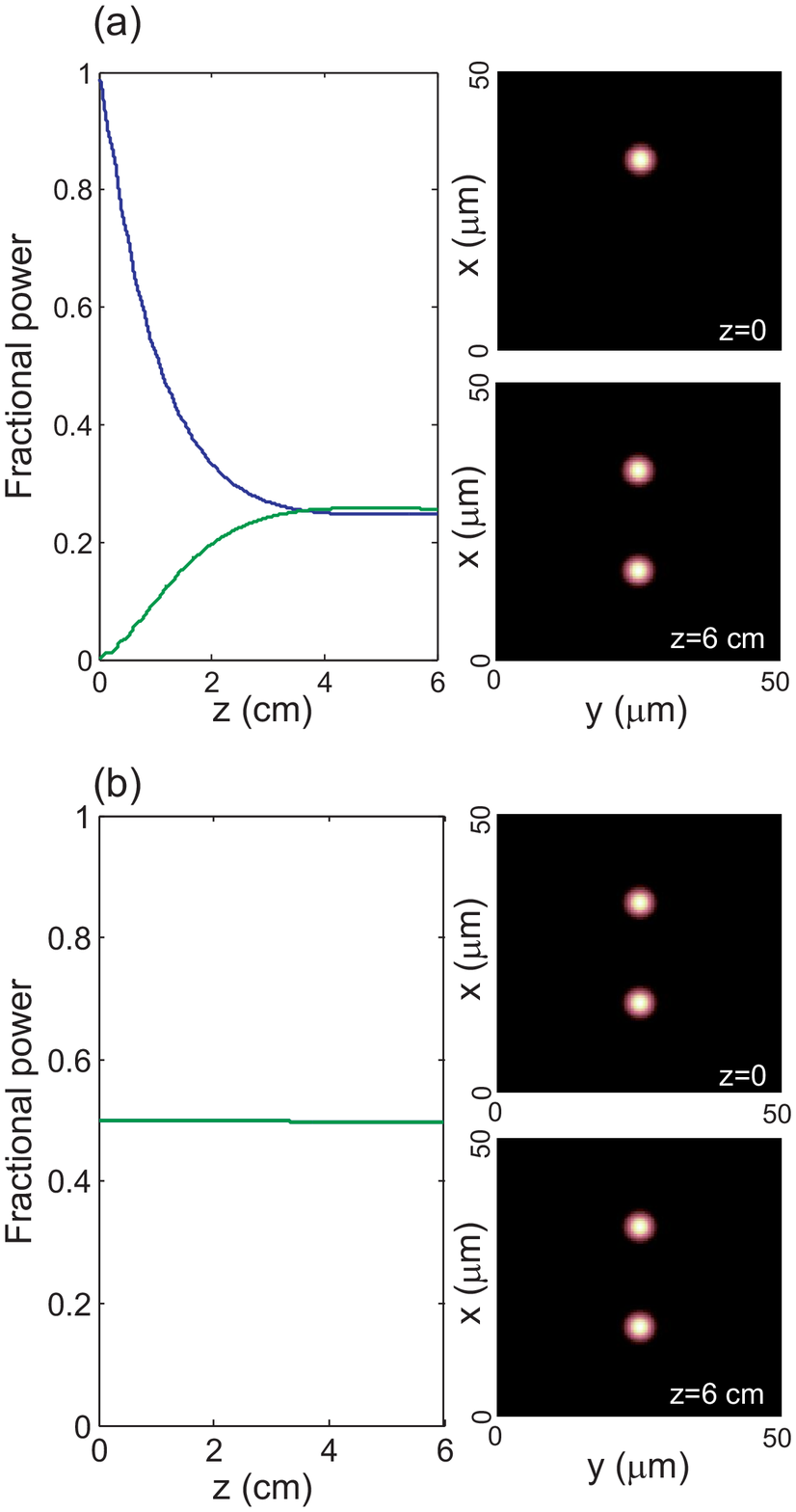}
\caption{(color online) Behavior of fractional light power trapped
in channel waveguides W$_1$ (blue curve) and W$_2$ (green curve)
versus propagation distance in a $L=6$-cm-long structure and
corresponding transverse light intensity distributions at the
input ($z=0$) and output ($z=6$ cm) planes, as obtained by
numerical simulations of the wave equation (1) for two different
excitation conditions. In (a) waveguide W$_1$ is excited in its
fundamental mode, leading to fractional decay. In (b) the trapped
state is excited, corresponding to full suppression of light
leakage into the slab S. Parameter values used in the simulations
are: $\lambda=980$ nm, $n_s=1.52$, $\Delta n_S=0.0049$, $\Delta
n_g=0.01$, $d=9 \; \mu$m, $r_c= 2 \; \mu$m, and $a=4 \; \mu$m.}
\end{figure}


\begin{thebibliography}{}
\bibitem{note1}
 In the Heisenberg picture, the state vector of the system would be
 constant whereas the
operator $\hat{\phi}$ evolve according to $i \hbar (d
\hat{\phi}/dz)=[\hat{\phi},\hat{H}]$. One can easily check that,
with the Hamiltonian operator given by Eq.(18), the evolution
equation of the operator $\hat{\phi}$ is precisely Eq.(3), where
the c-field $\phi$ is replaced by the operator $\hat{\phi}$.
\bibitem{note2}
More precisely, the monochromatic limit is attained by assuming in
Eqs.(21) and (22)
$f^{(n)}(\mathbf{q},\mathbf{\Omega},z)=g^{(n)}(\mathbf{\mathbf{q}},z)h(\Omega_1)h(\Omega_2)...h(\Omega_n)$,
where $h(\Omega)$ is a spectrally narrow function at around
$\Omega=0$ normalized such that $\int d \Omega |h(\Omega)|^2=1$,
and $g^{(n)}(\mathbf{q},z)$ satisfies Eq.(22) with $\Omega_l=0$.
The operator $\hat \phi(\rho)$ is then formally defined as $\hat
\phi(\rho)=\int d \Omega h(\Omega) \hat \phi(\rho,\Omega)$. The
$n$-photon number state $|f^{(n)}\rangle$, given by Eq.(21), takes
then the simplified form $|f^{(n)}\rangle=(1/(\sqrt{n!}) \int d
\mathbf{q} g^{(n)}(\mathbf{q},z) \hat \phi^{\dag}(\rho_1) \hat
\phi^{\dag}(\rho_2) ... \hat \phi^{\dag}(\rho_n)|0\rangle $.
\bibitem{note3}
As the waveguides W$_1$ and W$_2$ are symmetrically placed at
opposite sides from the slab waveguide S, if $(c_1(z), c_2(z),
c_k(z))$ is a solution to Eqs.(5-7) with initial condition
$c_1(0)=1$, $c_2(0)=0$ and $c_k(0)=0$, then $(c_2(z), c_1(z),
c_k(z))$ is the solution to Eqs.(5-7) with  reversed
 initial condition $c_1(0)=0$, $c_2(0)=1$ and $c_k(0)=0$. Light
tunnelled into the slab waveguide S is thus given by the same
superposition of continuous modes ($\sim \int dk c_k(z)
u_k(\rho)$) for the two different initial conditions. This means
that the last two terms on the right hand sides of Eqs.(38) and
(39) are equal each other.
\end{thebibliography}
\end{document}